# *Ab initio* study of the optical phonons in 1D antiferromagnet $Ca_2CuO_3$

Nam Nhat Hoang[a)], Thu Hang Nguyen, and Chau Nguyen

Center for Materials Science, Vietnam National University, 334 Nguyen Trai, Hanoi, Vietnam

In the spin 1/2 antiferromagnetic $Ca_2CuO_3$ the optical excitation along 1D a Cu-O chain showed the enrichment of forbidden peaks that could not be explained by the group theory. We present the cluster-model *ab initio* study of these optical phonons based on the Hartree-Fock SCF calculation with the 3-21G basic set. The obtained results showed very good agreement with the observed data. The Cu-O resonances generally showed the lower shifts in $Ca_2CuO_3$ than in pure CuO and were primarily composed of the vibrations of the oxygen in static host lattice whereas the Cu movements happened only in the collective lattice vibrations. Almost complete classification of the forbidden phonons is presented.

[a)]Electronic mail: namnhat@gmail.com

The importance of the low dimensional system $Ca_2CuO_3$ in both practical and fundamental aspects has attracted much attention from scientists during the last decades. This system exhibits various different properties associated with its low dimensionality, such as the covalent insulation[1], the van Hove singularity on the spin Fermi surface[2] and the spin-charge separation[3,4]. The structure of $Ca_2CuO_3$ (schematically featured in Fig. 1(a)) is very similar to the 2D superconducting $La_2CuO_4$: there is only oxygen lacking which perpendicularly connects two parallel Cu-O chains. Some compounds with the $Ca_2CuO_3$ structure, *e.g.* an oxygen excessive $Sr_2CuO_{3.1}$, can transform its structure under pressure into the $La_2CuO_4$ type structure and become a high $T_c$ superconductor ($T_c$=70K)[5]. The $Ca_2CuO_3$ exhibits a strong spin 1/2 antiferromagnetic coupling along its 1D Cu-O(2) chains. The intrachain exchange integral $J_\parallel \approx 0.6$eV, estimated on the basis of the *t-J* model, shows a record high value among the 1D systems and is about 300 times greater than the interchain coupling $J_\perp$[6–9]. With this observation, the structure of Raman-active phonons along the Cu-O(2) chain direction is enriched by features that are normally forbidden while in the other two directions only two $A_g$-mode phonons are visible. The first experimental study of the optical phonons has been presented by Yoshida *et al.*[10], Zlateva *et al.*[11] and later by Bobovich *et al.*[12] and Hoang *et al.*[13]. The first two studies reported the measurement on the single crystals, whereas the last two reported on the powder samples. Despite differences in chemical contents of the samples, which followed either from the differences in preparation routes or from the doping of further elements (*e.g.* Sr or U), the discussed phonon structures agreed quite well with each other. There are also two theoretical results available for the undoped $Ca_2CuO_3$: one is from the lattice dynamic calculation[11] and another from the tight-binding approach[14]. As these studies showed, there was a strong coupling between the forbidden phonons and the intrachain charge-transfer process mediated by the electrons excited by light. Although several observed features have their correct explanation, the problem still remains for the assignment of Cu-O resonances and the majority of overtones. It is also worth-while to mention that not all phonons can be classified as originating from the pure $Ca_2CuO_3$ phase. Recent study has showed that there was always a recognizable amount of the CuO phase presented in the final $Ca_2CuO_3$ samples which have been prepared by the ceramic technology[12,13,15,16].

For the pure and the Sr-doped, U-doped $Ca_2CuO_3$, several Raman studies are available[10–14]. Fig. 2 (upper part) shows the measured data using the light from He-Ne laser with $\lambda$ = 623.8 *nm* (*i.e.* 1.96 eV, note that the maximal scattering output occurs at 2.0 eV[10]). From this figure, the peaks are seen at 200, 280, 307, 467, 530, 663, 890, 942, 1142, 1217, 1337 $cm^{-1}$. This structure represents the most complete picture of all observed Raman-active optical phonons in the $Ca_2CuO_3$. For the scattering light from Nd:YAG laser ($\lambda$=1064*nm*, *i.e.* 1.17eV), some peaks disappeared (*i.e.* 200, 467 and 942 $cm^{-1}$) but the main features remained the same[12,13]. It is obvious that the structure of the Raman spectra depends on energy of the excitation light and for our case the He-Ne laser provided a more complete set of scattering lines. In Table I we summarized all observed frequencies. We now revise how these peaks have been assigned in Ref. 10 and 11. From the symmetry analysis, in the space group *Immm* ($D_{2h}^{25}$), the optical phonons at the Γ point ($\boldsymbol{k} = \boldsymbol{0}$) compose of 6 Raman active modes ($2A_g+2B_{1g}+2B_{2g}$) and 9 IR active modes ($3B_{1u}+3B_{2u}+3B_{3u}$). The $A_g$-, $B_{1g}$-, $B_{2g}$-mode phonons associate with the Wyckoff site *4f* (site symmetry $C_{2v}^2$) of the Ca and O(1), so with the vibrations of these atoms along axis *c* ($A_g$) and *a, b* ($B_{1g}, B_{2g}$). The $A_g$-mode phonons are active in the (*a, a*), (*b, b*) and (*c, c*) geometry and the $B_{1g}$-, $B_{2g}$-mode phonons are allowed only in the (*a, c*) and (*b, c*) settings. So by performing the scattering measurement in these exact configurations with some single crystal pieces, the $A_g$-, $B_{1g}$-, $B_{2g}$-mode phonons can be determined. Indeed, Yoshida *et al.*[10] has identified the $A_g$-mode phonons to be 306 $cm^{-1}$ (assigned to the Ca movement) and 530 $cm^{-1}$ (assigned to the O(1) movement). These two phonons were the sole phonons in the $c(a,a)\bar{c}$ and $a(c,c)\bar{a}$ configurations, so the assignments were unique. However, no structures due to the $B_{1g}$-, $B_{2g}$-mode phonons were experimentally observed in the respective scattering configurations[10, 11].

The rich features only appeared for the $a(b,b)\bar{a}$ configuration, *i.e.*

when the light polarization was parallel to axis *b*. Yoshida *et al.*[10] reported the following lines: 235, 306, 440, 500, 690, 880, 940, 1140, 1200 and 1330 $cm^{-1}$. All these peaks, except the one at 500 $cm^{-1}$ (not seen in Ref. 11 and 12), have their counterparts in the spectra in Fig. 2 (upper part). The weak features that were also visible (but not discussed) in Ref. 10 correspond closely to 200, 470, 640, 1000 and 1390 $cm^{-1}$; the first two of them were also reported in Ref. 11. This peak structure is richer than that offered by the symmetry analysis. Among them, the 440, 500 and 690 $cm^{-1}$ were ascribed as the first-order zone-boundary phonons (T-point with **k** = (0.5,0.5,0)), whereas the 880, 940, 1140, 1200 and 1330 $cm^{-1}$ as their high-order two phonon scattering[10]. Since the 440, 690 $cm^{-1}$ lines were also observed for both doped and undoped $Ca_2CuO_3$ (440, 670 $cm^{-1}$ in Ref. 12; 430, 690 $cm^{-1}$ in Ref. 11; 430, 670 in Ref. 13) Zlateva *et al.*[11] has suggested that all extra lines in the Raman spectra are due to the high-order scattering. This consideration resources in the finite and segmented Cu-O(2) chains of different lengths in the real polycrystalline samples, which leads expectedly to the overtones. It may however result from the impure phases presented as it was difficult to exclude all CuO, CaO and $CaCu_2O_3$ phases from the final product by means of the ceramic and oxalate co-precipitation techniques[15,16].

The $B_{1u}$-, $B_{2u}$-, $B_{3u}$-mode phonons, associated with all Wyckoff sites in the *Immm* space group (namely, *2d* of Cu, *2a* of O(2), *4f* of Ca and O(1)), correspond to the vibration of these atoms along the crystallographic axis *c, b* and *a* respectively. As these modes are IR active, they can be observed in the reflectivity measurement for light polarization along each axis[10] or in the IR transmission measurement[11]. The following lines were reported in Ref. 10 (TO-phonons): 215, 340, 660 $cm^{-1}$ ($B_{2u}$); 260, 410, 460, 580 $cm^{-1}$ ($B_{1u}$ and $B_{3u}$); the additional structures were found at 350 and 540 $cm^{-1}$ and were ascribed as the $B_{1u}$- and $B_{3u}$-mode phonons in Ref. 11. Most of these peaks are reproduced in Fig. 2 (lower part).

For the purpose of classification of all vibrational states, we performed the *ab initio* study on the model cluster illustrated in Fig. 1(b) with the Gaussian 2003 software[17]. This is a medium sized layer model stacking one Cu-O layer between the other two Ca-O layers. One of the difficulties with the cluster model, beside the usual convergence problems and vast computational costs, is that the symmetry of the local models is not the same as that of the real compound. This introduces several additional model-specific lines into the output spectra. Those 'phantom lines' can be partly identified by investigating various models of different shapes and sizes, but they cannot be avoided in principle. Six different clusters were involved in the calculation: (1) starting from the $Ca_4Cu_2O_8$ cluster by adding a unit $Ca_2Cu_2O_6$ to form the 2-fold, 3-fold structures $Ca_6Cu_4O_{14}$ and $Ca_8Cu_6O_{20}$; (2) starting from a 6-fold cluster $Ca_{18}Cu_8O_{28}$ (Fig. 1(b)) by adding a unit $Ca_6Cu_4O_{12}$ to form the 9-fold and 12-fold structures $Ca_{24}Cu_{12}O_{40}$ and $Ca_{30}Cu_{16}O_{52}$. The largest cluster contains 938 basis functions (molecular orbitals - MOs) for the UHF/STO-3G setting (746 paired electron occupied MOs and 192 unoccupied MOs). It is reasonable that the higher level theories can be used for the smaller clusters, such as the Density Functional Theory with some larger basic sets. But for the larger clusters (above 6-fold), the calculation was performed using the Self Consistent Field (SCF) Hartree-Fock (HF) method with the unrestricted spin model (UHF) on the 3-21G wave function basic set. The more compact restricted spin Hartree-Fock model (RHF) was successful in the so-called single point energy calculation (integral accuracy reduced to $10^{-5}$) but usually failed in the second derivatives calculation (when the integral accuracy increased to $10^{-8}$). For the smaller clusters, the stability tests showed that there was a transition from the RHF to UHF, *i.e.* the UHF wave functions usually provided the lower energy minimum. With the increase in cluster size, there was a considerable difference in the output spectra when the smaller STO-3G set was substituted for the 3-21G set. However, the difference was not large if the 6-31G set replaced the 3-21G set. It is preferably to chose the larger sets but for the relatively large sizes of the studied clusters, the 3-21G set provided optimal computational efficiency at present time. Larger settings, *e.g.* the DFT/6-31G required an extra amount of storage which exceeded the 4GB limit for the file size in most file systems. The frequency computation was accomplished with the Mulliken charge analysis and the thermochemistry analysis for the vibrational states.

Excluding the vibrations that are specifically associated with the atoms lying at the cluster boundary, the final calculated Raman and IR spectra are shown in Fig. 3. These spectra belong to the medium-size cluster $Ca_{18}Cu_8O_{28}$.

From the analysis of simulated vibrational states three IR-active $B_{2u}$ frequencies 210, 337, 657 $cm^{-1}$ correspond to the vibration of Cu, O(1) and O(2) along *b* axis. These lines have been assigned in Ref. 10 to the same atoms, however, the *ab initio* results shows some slight movement of Ca with the 210 $cm^{-1}$ line. The $B_{3u}$-phonon at 351 and the $B_{1u}$-phonons at 548, 589 $cm^{-1}$ associate with the vibration of O(2) along axis *a* and *c* respectively; the O(1) atoms also participate in the 351 line. Here again the assignment is the same as in Ref. 10. The other $B_{1u}$-phonon at 410 and $B_{3u}$-phonon at 457 $cm^{-1}$ originate in the moving of O(1) along *c* or *a*. In Ref. 10 and 11 the O(2) movement along axis *a* has been assigned to the 457 line. The rest peak, the $B_{1u}$-phonon seen at 265 $cm^{-1}$, follows from the breathing vibration involving both Cu and Ca transition along axis *a*. This peak has been considered as resulting from the sole movement of Cu in the previous studies[10,11].

The assignments for the two Raman-active $A_g$-mode phonons 306 and 528 $cm^{-1}$ are the same as in Ref. 10. These phonons are caused by the moving of the Ca and O(1) along axis *c* in nearly static host lattice.

Among the Raman-forbidden lines that were considered as the overtones in the previous studies[10, 11], the peaks at 211, 231 and 288 $cm^{-1}$ follow mainly from the movement of O(2) along axis *a* (288 line) plus the breathing vibration (211) or the movement of Ca in (*b*,*c*) plane (231).

The peaks 440 and 461 $cm^{-1}$ originate from the vibration of O(1) along *c* (440) plus O(2) along *b* (461). The shift at 512 $cm^{-1}$ (observed also in the Sr-doped $Ca_2CuO_3$[10,11]) is due to the displacement of both O(1) along axis *a* and O(2) in (*a,b*) plane. The sole O(2) stretching resonance along axis *b* is responsible for the 630 $cm^{-1}$ forbidden line. The illustration is given in Fig. 4 for the 211 and 512 $cm^{-1}$ lines.

It is worth noting that in $Ca_2CuO_3$ the Cu-O(2) resonances showed the lower frequencies in comparison with the Cu-O resonances in pure CuO, *e.g.* 288 *vs.* 298 $cm^{-1}$, 630 *vs.* 632 $cm^{-1}$. This agrees with the smaller force constant for the Cu-O bonding in $Ca_2CuO_3$ which is partly demonstrated by the longer average bond distance: 1.889Å in $Ca_2CuO_3$ vs. 1.875Å in CuO. From the charge analysis, the valence distributed within the Cu-O bonds in the pure CuO is also a little higher than in the $Ca_2CuO_3$.

For the shifts associated with the Ca-O resonances, two lines are seen at 231 and 1000 $cm^{-1}$. Although the 1000 $cm^{-1}$ peak is suggested as the two phonon scattering from the 500 $cm^{-1}$ line, there is no reason to exclude it from being considered as originating from the impure CaO.

For the Raman shifts which correspond to the vibration of the Cu, the *ab initio* results showed that there was no simple vibration of Cu in the static host lattice. All vibrations involving the Cu atoms are mainly the collective lattice vibrations in which the O(2) atoms participate (*e.g.* the 211 $cm^{-1}$ line). This observation agrees well with the structural analysis of rigidity of the Cu-O(2) bonds (axis *b*) previously given in Ref. 13, 15 and with the strong coupling of phonons in the 1D Cu-O(2) chain with electron-hole pairs created during excitation by light[10,14]. Such coupling is a very typical phenomenon in the superconducting cuprates. The doping in $Ca_2CuO_3$ seems to have only a little effect on its phonon structure as all known cases until now (*e.g.* Sr-doped[10,11] and U-doped[13]) did not show any divert features.

From the analysis given, the Cu-O(2) resonances in $Ca_2CuO_3$ are strongly coupled with the collective lattice breathing vibrations while most of the rest of the phonons originate from the sole vibrations of the oxygen in nearly static host lattice. For more accurate results, the Density Functional Theory calculation should be involved with some larger basic sets such as the 6-31G. Considering computational costs at the present time, we leave this for the future.

The authors would like to thank the Grant Projects No. QG-07-02 and DTCB 405 506 for the financial supports.

TABLE I. The Raman and IR frequencies ($cm^{-1}$) for $Ca_2CuO_3$. Comparisons are given to the pure[11,13], the Sr-doped $Ca_2CuO_3$[10,11] and to the theoretical values obtained by the lattice dynamic calculation[11] and the tight-binding approach[14]. For the Raman-forbidden lines, the values presented in parentheses correspond to the additional features visible in Fig. 4, Ref.10 but not reported by its authors.

| Assignment (BV=breathing vibration) | | Optical phonons in $Ca_2CuO_3$ | | | | | |
|---|---|---|---|---|---|---|---|
| | | Experimental | | | Theoretical | | |
| [10, 11] | *This work* | Ref.10 | Ref.11 | Ref.13 | Ref.11 | Ref.14 | *This work* |
| $A_g$-mode phonons (Raman active) (*c* axis) | | | | | | | |
| Ca | Ca | 306 | 311 | 307 | 311 | | 306 |
| O(1) | O(1) | 530 | 531 | 530 | 531 | 530 | 528 |
| $B_{2u}$-mode phonons (IR-active) (*b*-axis) | | | | | | | |
| Cu | Cu | 215 | 225 | 215 | 201 | | 210 |
| O(1) | O(1) | 340 | 354 | 350 | 371 | | 337 |
| O(2) | O(2) | 660 | 682 | 670 | 673 | 700 | 657 |
| $B_{1u}$- and $B_{3u}$-mode phonons (IR-active) (*c* and *a* axis) | | | | | | | |
| Cu ($B_{3u}$) | | | 194 | | 155 | 135 | |
| Cu ($B_{1u}$) | Cu,Ca∥a +BV ($B_{1u}$) | 260 | 278 | 272 | 291 | | 265 |
| O(1),O(2)($B_{3u}$) | O(2),O(1)∥a ($B_{3u}$) | 350 | 354 | 350 | 337 | | 351 |
| O(1) ($B_{1u}$) | O(1)∥c ($B_{1u}$) | 410 | 412 | 415 | 400 | | 410 |
| O(2) ($B_{3u}$) | O(1)∥a ($B_{3u}$) | 460 | 457 | 453 | 424 | 450 | 457 |
| O(2) ($B_{1u}$) | O(2)∥c ($B_{1u}$) | 540 | 530 | 532 | | | 548 |
| O(2) ($B_{1u}$) | O(2)∥c ($B_{1u}$) | 580 | | | 577 | | 589 |
| The Raman-forbidden lines (overtones) | | | | | | | |
| ? | O(2)∥a +BV | (200) | 203 | 200 | | | 211 |
| Cu | O(2)∥a +Ca∈(b,c) | 235 | | | | | 231 |
| ? | O(2)∥a | | 310 | 280 | | | 288 |
| T-point O(2) | O(1)∥c | 440 | 430 | | | 419 | 440 |
| 235+235 | O(1)∥c +O(2)∥b | (470) | 472 | 467 | | | 461 |
| O(1), O(2) | O(1)∥a +O(2)∈(a,b) | 500 | | | | 505 | 512 |
| ? | O(2)∥b +CuO? | (640) | | | | | 630 |
| O(1), O(2) | ? | 690 | 690 | 663 | | 670 | |
| 2 phonon | 440+440 | 880 | 880 | 890 | | | |
| 2 phonon | 440+500 | 940 | 940 | 942 | | | |
| 2 phonon | 500+500 or CaO? | (1000) | | | | | |
| 2 phonon | 440+690 | 1140 | | 1142 | | | |
| 2 phonon | 500+690 | 1200 | | 1217 | | | |
| 3 phonon | 440+440+440 | 1330 | | 1337 | | | |
| 2 phonon | 690+690 | (1390) | | | | | |

FIGURE CAPTIONS

FIG. 1. (Color online) The packing structure of 3 unit cells ($a \times 3b \times c$) for $Ca_2CuO_3$ (a) and the model cluster $Ca_{18}Cu_8O_{28}$ (b) used in the *ab initio* calculation of vibrational states.

FIG. 2. The Raman scattering (upper) and the FTIR transmission (lower) spectra of the pure $Ca_2CuO_3$. The Raman lines selected for listing in Table I are denoted by the arrows. The data for the graphs were taken from Ref. 13 with the permission from those authors.

FIG. 3. The simulated IR and Raman spectra for the $Ca_{18}Cu_8O_{28}$ cluster as obtained from the *ab initio* calculation using the Unrestricted spin Hartree-Fock SCF model with 3-21G basic set.

FIG. 4. (Color online) Two phases of the O(2) vibration along axis *a* in the forbidden 211 $cm^{-1}$ Raman shift (a) and the phases of the O(1) parallel movement along *a* together with the O(2) stretching motion in (*a,b*)-plane in 512 $cm^{-1}$ shift (b).

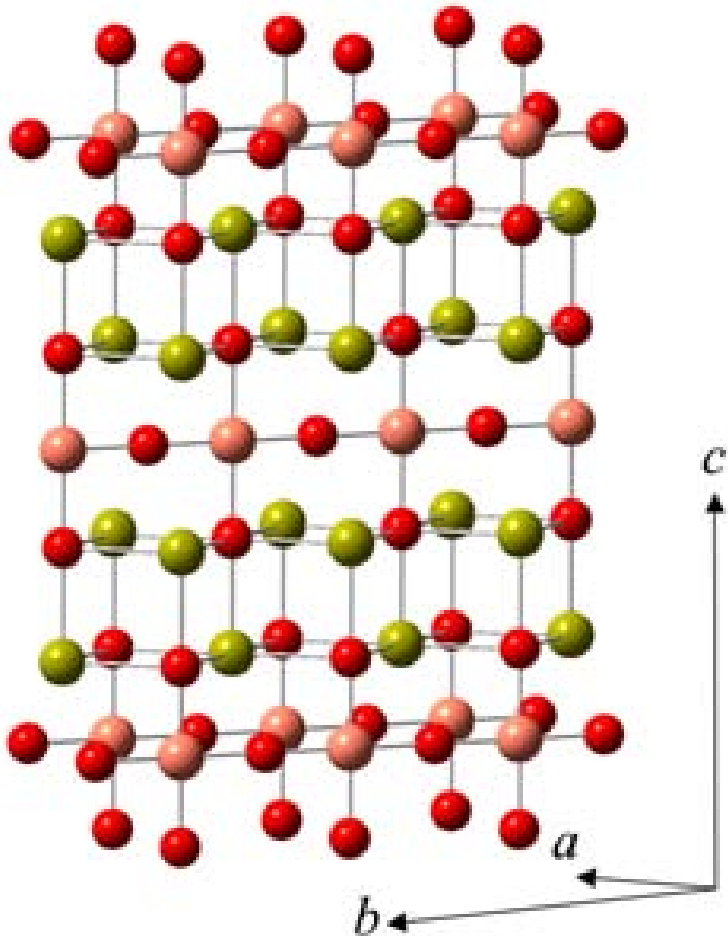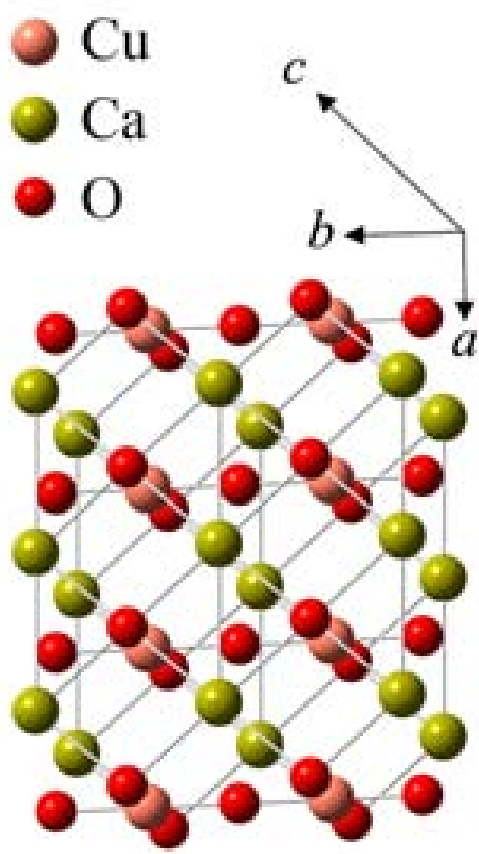

(a)　　　　　　　　　　(b)

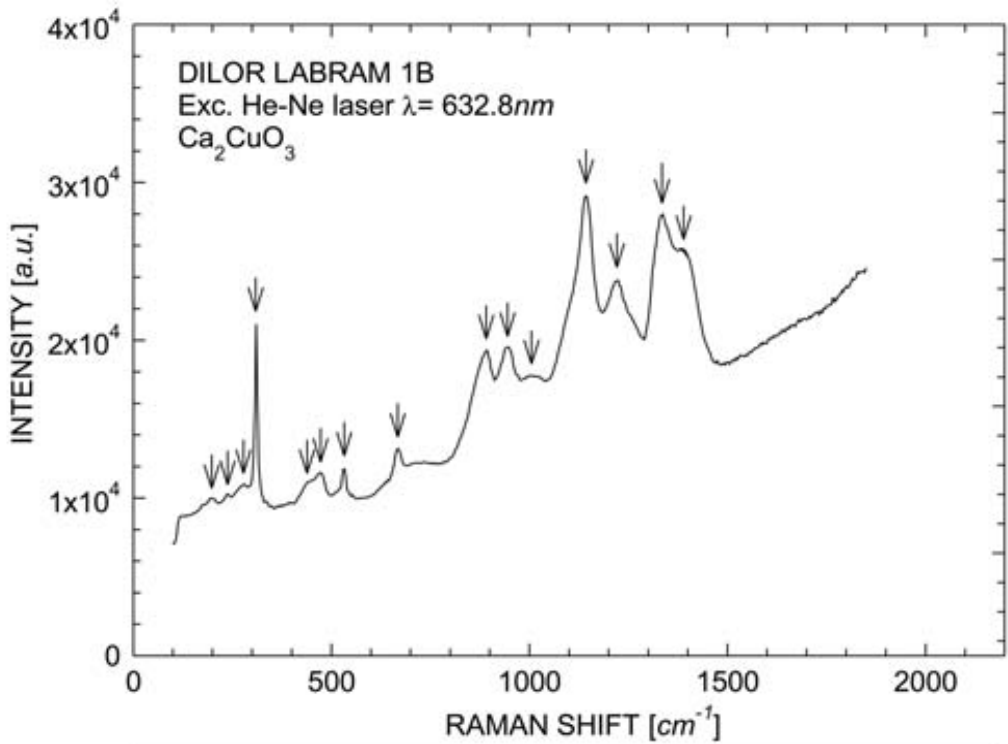

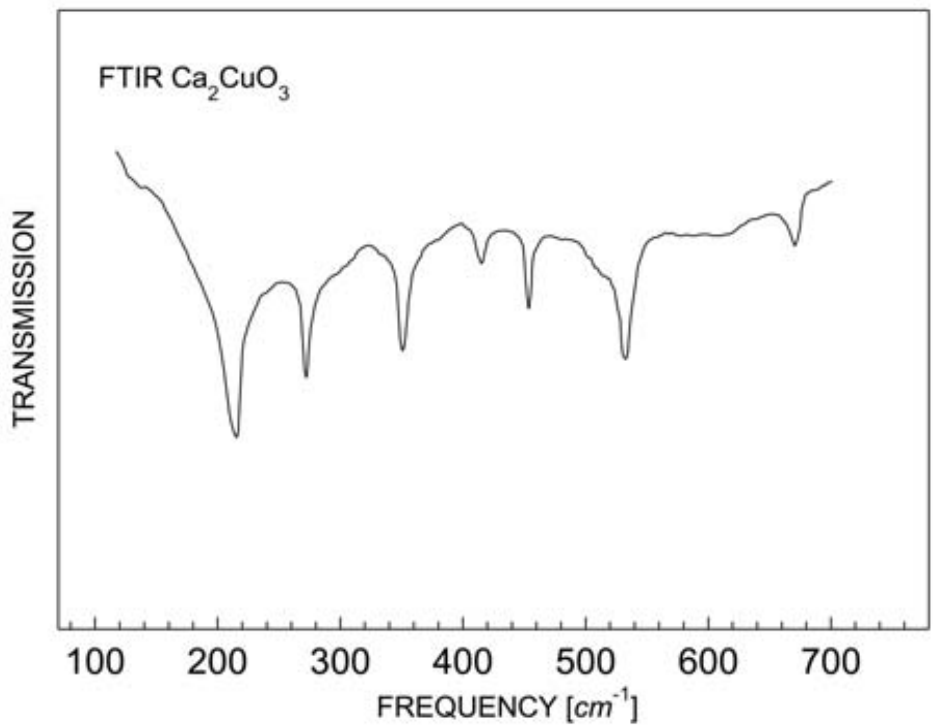

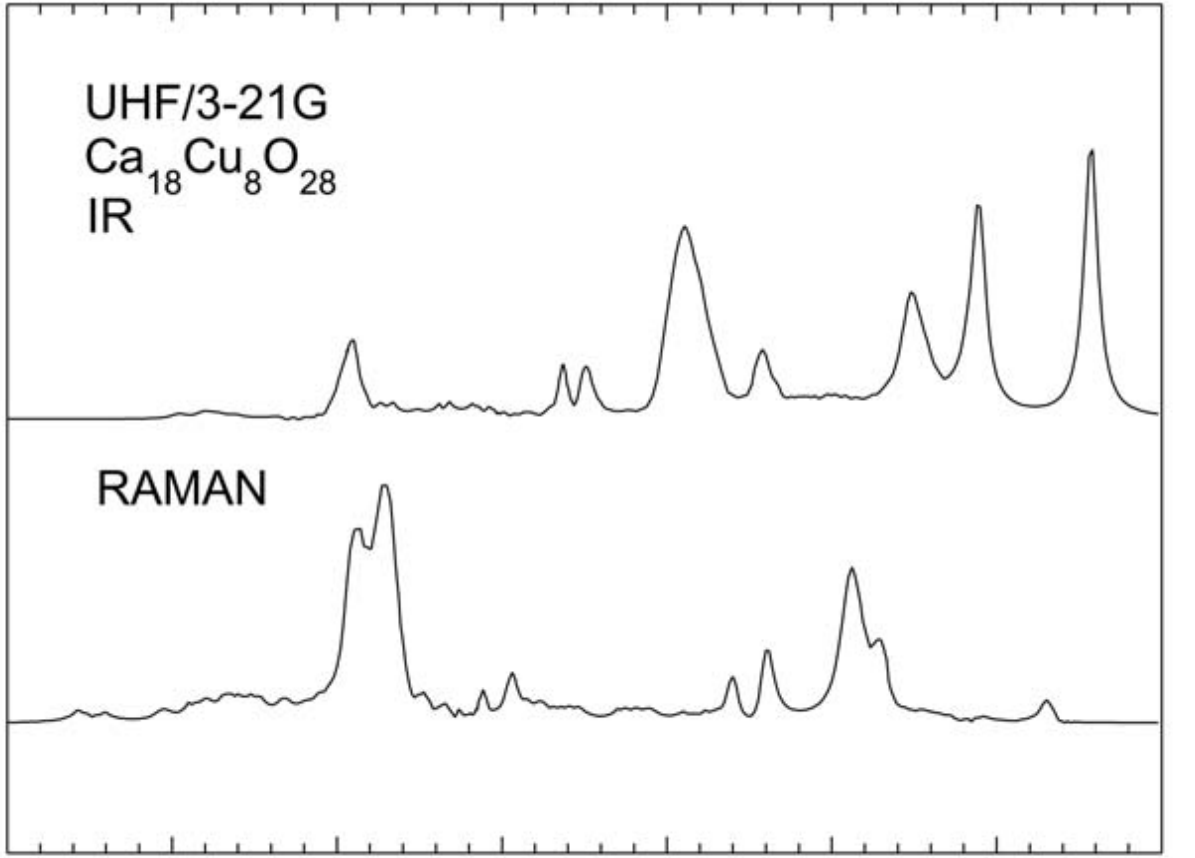

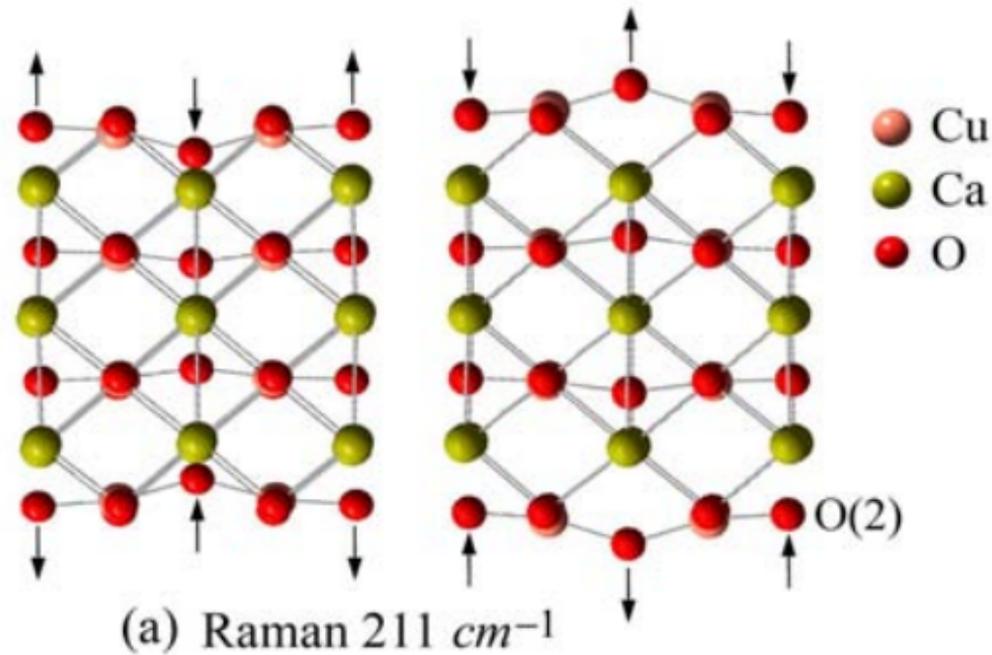

(a) Raman 211 $cm^{-1}$

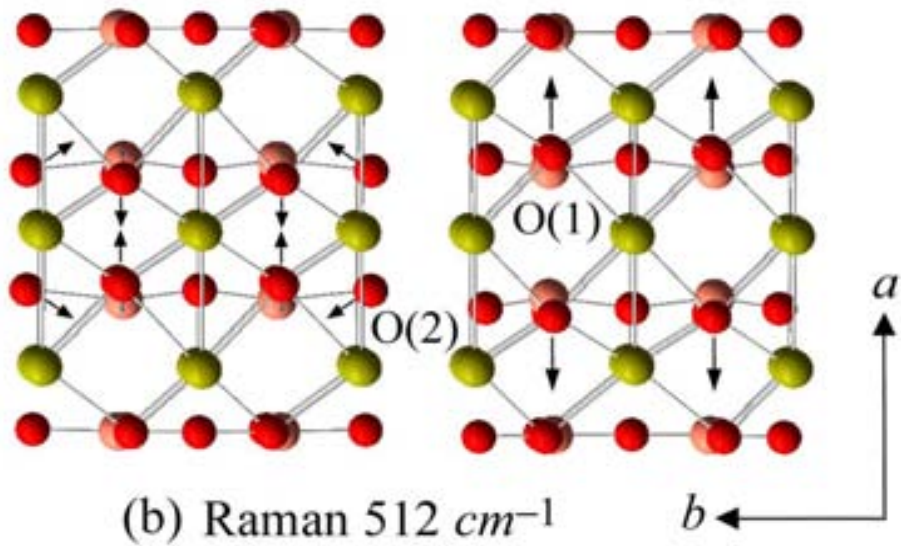

(b) Raman 512 $cm^{-1}$